\documentclass[letterpaper, 10 pt, conference]{ieeeconf}
\IEEEoverridecommandlockouts
 \usepackage{comment}

\usepackage{graphics} 
\usepackage{epsfig} 
\usepackage{times} 
\usepackage{amsmath} 
\usepackage{amssymb} 

\usepackage[english]{babel} 
\usepackage[ansinew]{inputenc}
\usepackage{mathtools}

\usepackage{todonotes}

\usepackage{lipsum} 

\usepackage[noadjust]{cite}

\title{\LARGE \bf Mixed Regular and Impulsive Sampled-data LQR} 

\author{Jamal Daafouz, J\'er\^ome Loh\'eac, Romain Postoyan
	\thanks{Work supported by grant OLYMPIA ANR-23-CE48-0006 and IUF.}
	\thanks{J. Daafouz, J. Loh\'eac and R. Postoyan are with the Universit\'e de Lorraine, CNRS, CRAN, F-54000 Nancy, France. 	{\tt\small firstname.name@univ-lorraine.fr} J. Daafouz is also with Institut Universitaire de France (IUF).}%
}

\newcommand{\R}{\mathbb{R}}
\newcommand{\N}{\mathbb{N}}
\newcommand{\Z}{\mathbb{Z}}
\newcommand{\Rlp}{\ensuremath{\mathbb{R}_{>0}}}
\newcommand{\Rlo}{\ensuremath{\mathbb{R}_{\geq 0}}}
\newcommand{\1}{{\bf 1}}
\newcommand{\0}{{\bf 0}}

\newtheorem{thm}{Theorem}

\newtheorem{defi}{Definition}

\newtheorem{rem}{Remark}

\newtheorem{lem}{Lemma}

\begin{document}

\maketitle
\thispagestyle{empty}
\pagestyle{empty}

\begin{abstract}
We investigate the benefits of combining regular and impulsive inputs for the control of sampled-data linear time-invariant systems. We first observe that adding an impulsive term to a regular, zero-order-hold controller  may help enlarging the set of sampling periods under which controllability is preserved by sampling. In this context, we provide a tailored Hautus-like necessary and sufficient condition under which  controllability of the mixed regular, impulsive (MRI)  sampled-data model is preserved. We then focus on LQR optimal control. After having presented the optimal controllers for the sampled-data LQR control in the MRI setting, we consider the scenario where an impulsive disturbance affects the dynamics and is known ahead of time. The solution to the so-called preview LQR is presented exploiting both regular and impulsive input components. 
Numerical examples, that include an insulin infusion benchmark, illustrate that leveraging both future disturbance information and MRI controls may lead to significant performance improvements.
\end{abstract}

\section{Introduction}

The challenge of designing control strategies that seamlessly integrate regular controls and impulses can be traced back to several decades ago. Already in 1971, R.~Bellman identified the potential of leveraging dynamic programming methods to generate both impulsive and regular controls for optimal drug administration \cite{Bellman1971}. In the conclusion of \cite{Bellman1971}, the need for future investigations combining impulsive and continuous treatments is acknowledged, but  few works have pursued this promising path, as far as we are aware of. 
A typical application involving a combination of regular and impulse controls is the basal bolus therapy for insulin administration in Type-1 Diabetes \cite{Reiterer19}. This strategy involves a continuous supply of a small amount of insulin called basal to maintain a relatively constant blood glucose level (regular control \cite{GOODWIN2015198}), with additional bolus insulin used to counterbalance disruptions (impulsive control~\cite{Good2019,Abuin19}). Nowadays, there is a resurgent interest in mixing regular and impulsive (MRI) control notably in the context of neuromorphic control, see, e.g., \cite{ribar2021neuromorphic}. However, rigorous methodological tools for optimal MRI control are scarce. In this work, we are interested  in MRI control for linear quadratic (LQ) problems for which few results are available as reviewed below.

In \cite{Motta96,Camilli1999}, optimal MRI control problems are formulated but for a specific class of nonlinear systems. These references delve into a rigorous mathematical analysis of optimal control behaviors, employing dynamic programming \cite{Motta96} and approximation methods \cite{Camilli1999}. In \cite{Aghdam06}, generalized sampled-data holds are investigated through the adaptation of basis functions to minimize a continuous-time LQR performance index. We can also cite e.g., 
\cite{Huang2012, Abuin19}. In~\cite{Huang2012}, the authors address the optimal LQR problem, where continuous control signals are combined with impulses at predefined times. The authors focus on finite-horizon costs while results for infinite-horizon costs are addressed only for purely impulsive controls. Finally, \cite{Abuin19} concentrates on accurately modeling systems controlled by pulses with a specified duration rather than impulses  (i.e., Dirac Delta function), which leads to a MPC strategy for insulin infusion control. On the other hand, and importantly, there is a natural connection between MRI control and impulsive, and more generally of hybrid dynamical systems, see, e.g., \cite{sanfelice-2021(hybrid-control-book)}, as impulsive inputs lead to state jumps. A relevant reference is thus  \cite{Teel20} where optimal hybrid LQ controllers are designed for linear time-invariant systems with periodic jumps.

In this paper and differently from the above references, we concentrate on the optimal \emph{sampled-data} control of linear time-invariant systems using MRI controls, where the regular term is held constant between two successive sampling instants. 
We first notice that adding an impulsive term to the input in this setting may only help enlarging the set of sampling periods under which controllability of the original continuous-time system is preserved. We then present a tailored necessary and sufficient condition for the controllability of the MRI sampled-data model, which simplifies classical Hautus test \cite{Hautus69} as it only involves a finite number of tests.  
We then focus on infinite-horizon sampled-data LQR control. It turns out that the optimal MRI-LQR control strategy is obtained by solving a standard Discrete Algebraic Riccati Equation (DARE). We then present what we believe is the main result of this work: preview LQR for sampled-data MRI models. By preview LQR \cite{Meinsma2006}, we mean to solve the LQR problem when the plant dynamics is affected by an impulsive disturbance, which is known ahead of time. This scenario is highly relevant in e.g., insulin infusion problem where meals can be considered as impulsive disturbances, which we may know relatively well in advance. Compared to e.g., \cite{Meinsma2006}, the originality here is to address sampled-data systems and MRI control. The relevance of MRI control for (preview) LQR control is illustrated via two examples. We first consider \cite[Example~1]{Souza2013}, for which the sampled-data model obtained with regular control leads to pathological sampling periods contrary to the MRI sampled-data model, whose controllability holds for any strictly positive sampling period. In this case, significant performance improvements are observed even at non-pathological sampling periods. We finally consider an insulin infusion problem, where the impulsive disturbance corresponds to a meal as mentioned above. In this case as well simulations show that the preview MRI-LQR controller is able to significantly reduce the peak of blood glucose level thereby confirming the benefits of MRI control for optimal drug administration as acknowledged in~\cite{Bellman1971}.

The rest of the paper is organized as follows. The MRI sampled-data model is derived in Section \ref{sect:problem-statement}. The controllability analysis is presented in Section \ref{sect:controllability}. Section~\ref{sect:lq-mri} is dedicated to (preview) LQR control using MRI control, whose results are illustrated on  examples in Section \ref{sect:example}. Section \ref{sect:conclusion} concludes the paper. The proof of Theorem~\ref{LQ-theorem-preview} is postponed to the appendix to avoid breaking the flow of exposition.\\


\noindent\textbf{Notation.} $\R$ stands for the set of real numbers, 
$\Rlo$ for the set of non-negative real numbers, $\Rlp:=\Rlo\backslash\{0\}$, $\Z$ for the set of relative integers, $\N$ for the set of non-negative integers and $\N_{>0}:=\N\backslash\{0\}$. Given $n,m\in\N_{>0}$, $\R^{n\times m}$ stands for the set of real matrices with $n$ rows and $m$ columns. We use $\delta$ to denote the Dirac delta function.
Given a matrix $M$, its null-space is denoted by $\text{ker}(M)$, its rank by $\text{rank}(M)$ and its spectrum by $\sigma(M)$ when $M$ is square.
Given square matrices $M_1,\ldots,M_n$ with $n\in\N_{>0}$, $\text{diag}(M_1,\ldots,M_n)$ is the block diagonal matrix whose block diagonal components are $M_1,\ldots,M_n$. We denote by $0_{n\times m}$ the zero matrix of $\R^{n\times m}$ with $n,m\in\N_{>0}$. For a complex number $z$, $\mathrm{Re}(z)$ and $\mathrm{Im}(z)$ stand for the real part and the imaginary part of $z$, respectively.

\section{Problem statement}\label{sect:problem-statement}

We consider a continuous-time linear dynamical system governed by the state space equation
\begin{equation}
	\dot{x}(t) = Ax(t) + B u(t), \quad x(0) = x_0,
	\label{eq:systemcon}
\end{equation}
where $x(t) \in \R^n$ is the state vector, $u(t) \in \R^m$ is the control input at time $t\in\R$, and $x_0\in\R^n$ is the initial condition with $n,m\in\N_{>0}$. We investigate the scenario where the input $u$ is generated using sampled data. In particular, we consider a periodic sequence of sampling instants $\{t_k\}_{k\in \N}$ verifying $t_k= kT$ for any $k\in\N$ with $T\in\Rlp$ the sampling period. Moreover, the input is composed of a regular term and of an impulsive term in the sense that 
\begin{equation}
	\left\{\begin{array}{r@{\,}c@{\,}ll}
        u(t) & = &  u_c(t) + u_i(t)       & \forall t\in\Rlo\\
        u_c(t) & = & u_{c_k}               & \forall t \in [t_k, t_{k+1}) \\
        u_i(t) & = & \delta(t-t_k)u_{i_k} & \forall t \in [t_k, t_{k+1})
\end{array}\right.	
 \label{eq:controlmri}
\end{equation}
where $u_c$ is a ``regular'' piecewise constant term with $u_{c_k}\in\R^{m}$, and $u_i$ is an impulsive term with $u_{i_k}\in\R^m$.

The exact discretized model of (\ref{eq:systemcon}) in closed-loop with (\ref{eq:controlmri}) at the sampling instants $t_k$, $k\in\N$, is given by
\begin{equation}
	{x}_{k+1} = A_{d,T}x_k + B_{d,T} u_{c_k} + B_{i,T}u_{i_k} 
	\label{eq:systemdis}
\end{equation}
where $x_k = x(t_k)$ and 
\begin{equation}
	\begin{array}{llll}\label{eq:sampled-matrices}
		\widetilde{A}_{T} \coloneqq \int_{0}^{T} e^{A\tau} d\tau, & A_{d,T} \coloneqq e^{AT},\\
		B_{d,T} \coloneqq {\widetilde{A}_{T}}B, & B_{i,T} \coloneqq A_{d,T}B.
	\end{array}
\end{equation}
We call (\ref{eq:systemdis}) the \emph{mixed regular and impulsive (MRI)} sampled-data model. 

The objective of this work is to demonstrate the benefits of combining regular and impulsive inputs as in~(\ref{eq:controlmri}) for the sampled-data control of~(\ref{eq:systemcon}). We first show in Section~\ref{sect:controllability} how adding an impulsive component to a regular  controller may help to ensure the controllability of system~(\ref{eq:systemdis}) for a larger class of sampling periods~$T$ than when just considering regular inputs. We then focus on the LQR  problem in Section~\ref{sect:lq-mri}. After having presented the solution to the LQR  control in the mixed regular impulsive control setting, we focus on preview LQR  control in the sense that future information on the disturbances is available. 

\section{Controllability analysis}\label{sect:controllability}

When the control input is regular, and does not consist of an additional impulsive term, i.e., $u_{i_k}=0$ for any $k\in\N$ in~(\ref{eq:controlmri}), we recover the classical linear sampled-data equation 
\begin{equation}
	{x}_{k+1} = A_{d,T}x_k + B_{d,T} u_{c_k}.
	\label{eq:systemstand}
\end{equation}
We first observe a major difference between (\ref{eq:systemdis}) and (\ref{eq:systemstand}): the input matrix of the MRI sampled-data model in (\ref{eq:systemdis}), namely $[B_{d,T}\,\,B_{i,T}]$, is made of extra columns compared to the input matrix of the standard model (\ref{eq:systemstand}), namely $B_{d,T}$. Since both of these models have the same state matrix $A_{d,T}$, adding an impulsive input may only help enlarging the set of sampling periods under which controllability is preserved by sampling. To highlight this difference, 
we consider \cite[Example~1]{Souza2013} that is, 
\begin{equation}
A = \left[ \begin{array}{cc}
		0 & 1\\ -6 & 1
\end{array} \right] , \quad
B = \left[\begin{array}{c} 0 \\ 1 \end{array}\right].
\label{eq:modAB}
\end{equation}
The continuous-time system (\ref{eq:systemcon}) with (\ref{eq:modAB}) is controllable.
However, this controllability property may be lost by sampling due to the so-called pathological sampling.

\begin{defi}\label{def:pathological-sampling}
Suppose $(A,B)$ is controllable, $T\in\Rlp$ is a \emph{pathological sampling period for controllability} for system~(\ref{eq:systemstand}) (respectively, for system (\ref{eq:systemdis})) if $(A_T,B_{d,T})$ (respectively, $(A_T,[B_{d,T}\,\,B_{i,T}])$) is not controllable.	\hfill $\Box$
\end{defi}

Definition \ref{def:pathological-sampling} is related but differs to the notion of pathological sampling period in \cite[Chapter~3]{Chen1995}. In particular, any pathological sampling period for controllability as in Definition \ref{def:pathological-sampling} is a pathological sampling period in the sense of \cite[Chapter~3]{Chen1995}, but the opposite is not necessarily true as known in the literature, see, e.g., \cite{Kreisselmeier1999}. In the following, when we refer to a (non-)pathological sampling period for a given sampled-data system, we mean a (non-)pathological sampling period for controllability for this sampled-data system as in Definition \ref{def:pathological-sampling}. It appears that the pathological notions of \cite{Chen1995} coincide with the pathological notion (in the sense of Definition~\ref{def:pathological-sampling}) for system~(\ref{eq:systemstand}) for all the considered examples in this work.

With (\ref{eq:modAB}),  
the pathological sampling periods for system~(\ref{eq:systemstand}) are given~by
\begin{equation}\label{eq:pathological-T-first-example}
T = \frac{2\ell\pi}{\sqrt{23}}, \quad \ell \in\N_{>0}.
\end{equation}
For $T=\dfrac{2\ell\pi}{\sqrt{23}}$, we have
\begin{equation}\label{eq:matrices-insulin}
A_{d,T} = (-1)^le^{T/2}\begin{bmatrix} 1 & 0 \\
0 & 1 \end{bmatrix},\, 
B_{d,T} = \frac{1 - (-1)^\ell e^{T/2}}{6}\begin{bmatrix}
		1 \\0
\end{bmatrix}
\end{equation}
and the standard sampled-data model (\ref{eq:systemstand}) is not controllable. 
The MRI sampled-data model (\ref{eq:systemdis}), on the other hand, is characterized by the same matrices $A_{d,T}$, $B_{d,T}$ and
\begin{equation}\label{eq:Bi-insulin}
B_{i,T}=A_{d,T}B = (-1)^{\ell}e^{T/2}\begin{bmatrix}
		0 \\1
\end{bmatrix}
\end{equation}
and it is controllable for any $T$ verifying (\ref{eq:pathological-T-first-example}) and thus for any $T\in\Rlp$.
Therefore, for this example, adding an impulsive component to the control input helps to completely rule out  sampling periods under which controllability is lost by periodic sampling.

Nevertheless, this is not always the case, as exemplified next
\begin{equation}\label{eq:A-B-example-pathological-period-for-both-standard-and-MRI}
A=\begin{bmatrix}
    0 & -1\\
    1 & 0\\
\end{bmatrix},\quad B=\begin{bmatrix}
    0\\1
\end{bmatrix}.
\end{equation}
The continuous-time system (\ref{eq:systemcon}) with (\ref{eq:A-B-example-pathological-period-for-both-standard-and-MRI}) is controllable.
The eigenvalues of $A$ are $\pm i$. 
The corresponding pathological sampling periods for system~(\ref{eq:systemstand}) are given~by
$$T=\ell\pi, \quad \ell>1.$$
By setting $T=2\pi$, we have $A_{d,T}=\1$, $\widetilde{A}_T=0_{2\times2}$, $B_{d,T}=0_{2\times1}$ and $B_{i,T}=B$, and the pair $(A_{d,T},\begin{bmatrix}B_{d,T}&B_{i,T}\end{bmatrix})$ is not controllable.
That is to say, both the classical sampled-data model and the sampled-data MRI model are not controllable. As a consequence, MRI control does not always allow ruling out pathological sampling of the standard sampled-data model (\ref{eq:systemstand}). 

The next theorem proposes a reduced Hautus test to check the controllability of the MRI system~(\ref{eq:systemdis}). 
\begin{thm}\label{thm:controllability}
    Suppose the pair $(A,B)$ is controllable, and let $T\in\Rlp$.
    The pair $(A_{d,T},\begin{bmatrix}B_{d,T}&B_{i,T}\end{bmatrix})$ is controllable if and only if
    $$\ker\begin{bmatrix}
        A_{d,T}^\top-e^{\mu T} \1\\
        (\widetilde{A}_{T}B)^\top\\
        B^\top
    \end{bmatrix}=\{0\},$$
    for every $\mu\in\{\lambda\in\sigma(A) \text{ s.t. } \exists(\ell,\gamma)\in\Z\times(\sigma(A)\setminus\{\lambda\}) \text{ s.t. } (\lambda-\gamma)T=2i\pi\ell\}$.
    \strut\hfill$\Box$
\end{thm}

Theorem \ref{thm:controllability} is of interest when $T>0$ is a pathological sampling period for system~(\ref{eq:systemstand}). We say that this test is a reduced Hautus test, since, roughly speaking, the classical Hautus test has to be done on a reduced set of eigenvalues of~$A$. In fact, when $T>0$ is such that $T\neq 2i\pi\ell(\lambda_2-\lambda_1)$ for every $\ell\in\N$ and every $\lambda_1,\lambda_2\in\sigma(A)$, the pair $(A_{d,T},\begin{bmatrix}B_{d,T}&B_{i,T}\end{bmatrix})$ is controllable.
This fact follows from \cite[Theorem~3.2.1]{Chen1995}, since, according to this result, the pair $(A_{d,T},B_{d,T})$ is controllable. 
When $T>0$ is such that $T=2i\pi\ell(\lambda_2-\lambda_1)$ for some $\ell\in\N$ and $\lambda_1,\lambda_2\in\sigma(A)$, the addition of impulsive term can help to gain controllability of the system, this is for instance the case for the pair $(A,B)$ given by~(\ref{eq:modAB}).  
However, adding the impulsive input term does not always solve the controllability issue for the pair $(A_{d,T},\begin{bmatrix}B_{d,T}&B_{i,T}\end{bmatrix})$, see the example with matrices $A$ and $B$ given by~(\ref{eq:A-B-example-pathological-period-for-both-standard-and-MRI}). On the other hand, when $\ker(B^\top \widetilde{A}_T^\top)\cap \ker(B^\top)=\{0\}$, then the MRI system~(\ref{eq:systemdis}) is controllable. This is for instance the case for the pair $(A,B)$ given by~(\ref{eq:modAB}). 

\begin{proof}
    Let us first observe that $\sigma(A_{d,T})=\{e^{\lambda T},\ \lambda\in\sigma(A)\}$.
    Using Hautus test, we know that $(A_{d,T},B_{di,T})$ is controllable if and only if
    $$\ker(A_{d,T}^\top-e^{\lambda T} \1)\cap\ker(B^\top\widetilde{A}_{T}^\top)\cap\ker(B^\top A_{d,T}^\top)=\{0\},$$
    for every $\lambda\in\sigma(A)$.
    Since $A_{d,T}=e^{TA}$ is regular, this is equivalent to,
    $$\ker(A_{d,T}^\top-e^{\lambda T} \1)\cap\ker(B^\top\widetilde{A}_{T}^\top)\cap\ker(B^\top)=\{0\}.$$
    For every $\lambda\in\sigma(A)$ such that $e^{\lambda T}\not\in\{e^{\gamma T},\ \gamma\in\sigma(A)\setminus\{\lambda\}\}$, we have $\ker(A_{d,T}^\top-e^{\lambda T}\1)=\ker(A^\top-\lambda\1)$.
    But since the pair $(A,B)$ is controllable, we have $\ker(A^\top-\lambda\1)\cap \ker(B^\top)=\{0\}$.
    Hence, one just has to do the Hautus test for $\lambda\in\sigma(A)$ such that there exists $\gamma\in\sigma(A)\setminus\{\lambda\}$ such that $e^{\lambda T}=e^{\gamma T}$, that is to say that $\mathrm{Re}(\gamma)=\mathrm{Re}(\lambda)$, and $\mathrm{Im}(\lambda-\gamma)T=2\pi\ell$, for some $\ell\in\mathbb{Z}$.
\end{proof}
\begin{rem}\label{rk:finitHtest}
    It is important to observe that we only need a finite number of Hautus tests to guarantee the absence of pathological sampling periods for~(\ref{eq:systemdis}) or~(\ref{eq:systemstand}) according to Theorem \ref{thm:controllability}.
    In fact, $T>0$ is a potential pathological sampling period if there exists $a+ib_1, a+ib_2\in\sigma(A)\setminus\R$ such that $b_1\neq b_2$ and $T(b_2-b_1)=2\ell\pi $ for some $\ell\in\mathbb{Z}$, that is to say that $T=\ell T_0$, with $T_0=2\pi/(b_2-b_1)$.
    For every $\ell\in\mathbb{Z}$, we have
    \begin{align*}
        \int_0^{\ell T_0} e^{t(a+ib_1)} dt & = \frac{e^{\ell T_0(a+ib_1)}-1}{a+ib_1},\\
        \int_0^{\ell T_0} e^{t(a+ib_2)} dt & = \frac{e^{\ell T_0(a+ib_2)}-1}{a+ib_2}.
    \end{align*}
    Since $e^{iT_0b_1}=e^{iT_0b_2}$ (and hence $e^{i\ell T_0b_1}=e^{i\ell T_0b_2}$), we see that the coefficient at the numerator is the same for the two expressions.
    We also see that the coefficient at the denominator is independent of~$\ell$.
    Roughly speaking, this will ensure that the Hautus test made for the sampling time $\ell T_0$ will be equivalent to the one made for the sampling time~$T_0$.
    This is true unless $e^{\ell T_0(a+ib_1)}-1=0$, that is to say $a=0$ and $\ell T_0 b_1=2m\pi$ for some $m\in\Z$.
    Taking in account the expression of $T_0$, this means that $b_2/b_1$ is a rational number.\\
    In any cases, to ensure the controllability of the discrete system for every sampling period, the reduced Hautus test of Theorem~\ref{thm:controllability} has to be done for a finite number of (well-chosen) sampling periods~$T$.\\
    Note that, we have detailed here the argument for only two eigenvalues in $\sigma(A)\setminus\R$, but the general case can be treated similarly, i.e., by considering a (finite) sequence $a+ib_k\in\sigma(A)\setminus\R$ such that the $b_k$'s are two by two distinct and for $k>1$, $T(b_k-b_1)=2\ell_k\pi b$ with some $\ell_k\in\mathbb{Z}$.
    \strut\hfill $\Box$
\end{rem}
\begin{rem}
    Following the argument used in Remark~\ref{rk:finitHtest}, we can also state that if there exists $b\in\R\setminus\{0\}$ such that $ib\in\sigma(A)$, then there always exist a pathological sampling period for the system~(\ref{eq:systemstand}).
    More precisely, $T=2\pi/|b|$ is one of them.
    This is typically the case for the pair $(A,B)$ given by~\eqref{eq:A-B-example-pathological-period-for-both-standard-and-MRI}, where we have $b=1$.
    \strut\hfill $\Box$
\end{rem}
\section{Optimal MRI-LQR design}\label{sect:lq-mri}

Now that we have seen the possible benefits of MRI control for the controllability of sampled-data linear systems, we concentrate on the LQR  problem with MRI control. The cost function for system (\ref{eq:systemcon}), (\ref{eq:controlmri}) is introduced in Section \ref{subsect:cost} where we establish an equivalent expression along solutions to the sampled-data MRI model (\ref{eq:systemdis}). We then briefly present the corresponding optimal controllers in Section \ref{subsect:lq}. Afterwards, we state the main result of this section, namely preview optimal H$_{2}$ using MRI control, whose benefits are illustrated on numerical examples in Section \ref{sect:example}.


\subsection{Cost function}\label{subsect:cost}

Let $Q\in\R^{n\times n}$ be positive semi-definite and $R_c\in\R^{m\times m}$, $R_i\in\R^{m\times m}$ be positive definite. We define the next cost function, for any $x_0\in\R^n$ and $u=u_c+u_i$ with $u_c$, $u_i$ as in (\ref{eq:controlmri}),
\begin{equation}
\begin{array}{r@{\,}l}
	J(x_0,u) & := \displaystyle\int_0^{\infty} \Big( x(t)^\top Qx(t) + u_c(t)^\top R_c u_c(t) \\
 & \hspace{1.4cm}+ u_i^\top(t) R_iu_i(t)\Big) dt,
	\label{eq:critere} 
\end{array}
\end{equation}
where $x(t)$ is the solution to (\ref{eq:systemcon}) at time $t\in\Rlo$, initialized at $x_0$ with inputs $u=u_c+u_i$. The integral part related to the impulsive input has to be interpreted as $\sum_{k=0}^\infty u_{i_k}^\top R_i u_{i_k}$. 

The next lemma is a generalisation of \cite[Lemma 1]{Souza2013} to sampled-data MRI control. It provides an equivalent expression of cost (\ref{eq:critere}) considering solutions to the MRI sampled-data model (\ref{eq:systemdis}), thanks to the specific class of inputs in (\ref{eq:controlmri}). 

\begin{lem}\label{lem:equality-cost}
For any $x_0\in\R^n$, $u=u_c+u_i$ with $u_c$, $u_i$ in~(\ref{eq:controlmri}), 
\begin{equation}
J(x_0,u) = \displaystyle\sum_{k=0}^{\infty} \Big( x_k^\top Q_{d,T} x_k + 2 x_k^\top S_{d,T} v_k + v_k^\top R_{d,T} v_k \Big), 
\label{equalityJ}
\end{equation} 
where $x_k$ is the solution to (\ref{eq:systemdis}) at time $k\in\N$, initialized at $x_0$ with inputs $v_k=\left[\begin{matrix} u_{c_k} \\ u_{i_k} \end{matrix}\right]$ and 
\begin{equation}
\begin{array}{r@{\,}l}
Q_{d,T} & := \int_0^T (e^{As})^\top Q e^{As}ds \\
S_{d,T} & := \int_0^T (e^{As})^\top Q \widetilde{B}_{di}(s) ds \\
R_{d,T} & := \int_0^T \widetilde{B}_{di}(s)^\top Q \widetilde{B}_{di}(s)ds + \text{diag}(T R_c,R_i) \\
\widetilde{B}_{di}(s) & :=\begin{bmatrix} \int_0^{s} e^{A\tau}d\tau B, & e^{As}B\end{bmatrix}\quad\forall s\in\Rlo.
\end{array}
\label{notQRS}
\end{equation} \hfill $\Box$
\end{lem}
\begin{proof} Let $x_0\in\R^n$ and $u=u_c+u_i$ as in (\ref{eq:controlmri}). Cost~(\ref{eq:critere}) verifies
{\setlength\arraycolsep{0.5pt}%
\begin{multline}
	J(x_0,u)
 = \displaystyle\sum_{k=0}^{\infty}\! \int_{kT}^{{(k+1)}T}\!\!\! \Big( x(t)^\top Qx(t) + u_c(t)^\top R_c u_c(t) \Big) dt\\[-2mm]
 + \displaystyle\sum_{k=0}^{\infty}  u_{i_k}^\top R_i u_{i_k},
 		\label{eq:criterediscret1}
 	\end{multline}}
\hspace{-0.15cm}Using the change of variable $s=t-kT$ and as $u$ is given by (\ref{eq:controlmri}), we derive
\begin{equation}
\begin{array}{r@{\,}l}
	J(x_0,u) & = \displaystyle\sum_{k=0}^{\infty} \Big(\displaystyle\int_{0}^{T} x(s+kT)^\top Qx(s+kT) ds \\
 & \hspace{1.4cm}+ Tu_{c_k}^\top R_cu_{c_k} + u_{i_k}^\top R_iu_{i_k} \Big).
 \end{array}
\label{eq:criterediscret2}
\end{equation}
The desired result is obtained using (\ref{eq:systemcon})--(\ref{eq:systemdis}) and~(\ref{notQRS}); in particular, $x(s+kT)=e^{sA}x_k+e^{sA}Bu_{i_k}+\int_0^s e^{\tau A} d\tau Bu_{c_k}$.
\end{proof}



\subsection{Optimal design}\label{subsect:lq}

The next theorem provides the expressions of the optimal inputs minimizing (\ref{equalityJ}) along solutions to (\ref{eq:systemcon}), (\ref{eq:controlmri}). It is a direct application of \cite{shaiju08} in view of Lemma \ref{lem:equality-cost}. 
\begin{thm}\label{LQ-theorem}
Consider system (\ref{eq:systemcon}), (\ref{eq:controlmri}) with $(A,B)$ controllable, and $T\in\Rlp$ not pathological for system (\ref{eq:systemdis}). Given any $x\in\R^n$, the inputs minimizing (\ref{eq:critere}) are given by $u_{c_k}=K_cx_k$ and $u_{i_k}=K_i x_k$ for any $k\in\N$, where $x_k$ is the corresponding solution to (\ref{eq:systemdis}) initialized at $x_0$ and
\begin{equation}	
\left[\begin{array}{c} K_c \\ K_i	\end{array} \right] =	-(R_{d,T} + B_{di,T}^\top PB_{di,T})^{-1} (B_{di,T}^\top PA_{d,T} + S_{d,T}^\top)
\label{KcKiopt}
\end{equation}
with $B_{di,T}=\big[B_{d,T}~~B_{i,T}\big]$ and	$P=P^\top>0$ is solution to the DARE 
\begin{equation}
		\begin{array}{r@{\,}l}
			Q_{d,T} = & (A_{d,T}^\top PB_{di,T}+S_{d,T})(B_{di,T}^\top PB_{di,T}+R_{d,T})^{-1} \\
   & \times (A_{d,T}^\top PB_{di,T}+S_{d,T})^\top
			-A_{d,T}^\top PA_{d,T}+P.
		\end{array}
		\label{DARE1}
	\end{equation}
 \hfill $\Box$
\end{thm}


\subsection{Preview in H$_2$ optimal control}\label{subsect:preview}
Consider the dynamical system
\begin{equation}
\begin{array}{r@{\,}c@{\,}l}
\dot{x}(t) & = & Ax(t) + B(u_c(t) +  u_i(t))  + \widetilde{B} w(t), \,\, x(0) = 0 		\label{eq:system-w} \\
	z & = & Cx+D_cu_c + D_iu_i,
\end{array}
\end{equation}
with $u_c$, $u_i$ as in (\ref{eq:controlmri}), $z\in\R^p$,  $p\in\N_{>0}$, $C^\top=[Q^{1/2} ~~0_{n\times 2m}]^\top$, $D_c^\top=[0_{m\times p} ~~R_c^{1/2}~~0_{m\times p}]$,  $D_i^\top=[0_{2m\times p} ~~R_i^{1/2}]$ and $\widetilde B\in\R^n$.

Given $N\in\N_{>0}$ and $T\in\Rlp$, we focus on the optimal disturbance rejection of system (\ref{eq:system-w}) using sampled-data control of the form of (\ref{eq:controlmri}) when system (\ref{eq:system-w}) is affected by an impulsive disturbance $w(t)=\delta(t-NT)$ for any $t\in\Rlo$. Furthermore, we assume that $w$ is known ahead of time as in preview LQR \cite{Meinsma2006}. As it is usually done in the definition of the H$_2$-norm \cite[Chapter 6]{Chen1995}, the performance index is expressed as
\begin{equation}
	\widetilde{J}(u,w) := \displaystyle \int_0^\infty z (t)^\top z(t) dt.
	\label{CostH2}
\end{equation}
The optimal controller to the MRI-H$_2$ control problem with preview is given in the next theorem. This result extends \cite[Theorem 4]{Meinsma2006} to the MRI sampled-data case. Its proof is given in the appendix. 

\begin{thm}\label{LQ-theorem-preview}
Consider system (\ref{eq:system-w}), (\ref{eq:controlmri}) with $(A,B)$ controllable. Given $\widetilde B\in\R^n$, $T\in\Rlp$ not pathological for system (\ref{eq:systemdis}) and $N\in\N_{>0}$ such that $w(t) = \delta(t-NT)$ is known to the controller $NT$ units of time in advance. The inputs minimizing (\ref{CostH2}) are given by 
{\setlength\arraycolsep{0.5pt}\begin{equation}
\begin{array}{r@{\,}l}
\left[\begin{array}{c} u_{c_k} \\ u_{i_k} \end{array}\right] = & -(R_{d,T}\!+\!B_{di,T}^\top PB_{di,T})^{-1} (B_{di,T}^\top PA_{d,T}\!+\!S_{d,T}^\top )x_k \\
& -(R_{d,T}\!+\!B_{di,T}^\top PB_{di,T})^{-1}B_{di,T}^\top (G^\top )^{N-k-1}P\widetilde{B}
\end{array}
\label{u1et}
\end{equation}}
\hspace{-0.15cm}for $k \in \{0, \ldots, N-1\}$ and by 
{\setlength\arraycolsep{0.5pt}\begin{equation}
\left[\begin{array}{c} u_{c_k} \\ u_{i_k} \end{array}\right] = -(R_{d,T}+B_{di,T}^\top PB_{di,T})^{-1}(B_{di,T}^\top PA_{d,T}+S_{d,T}^\top )x_k\label{u2et}
\end{equation}}
\hspace{-0.15cm}for $k\geq N$, where $x_k$ is the corresponding solution to (\ref{eq:systemdis}) at time $k$,  initialized at $\widetilde B$, and $P=P^\top >0$ is solution to (\ref{DARE1}). Moreover, the optimal cost denoted $\widetilde J^\star$ is given by
\begin{equation}
\widetilde{J}^\star(\widetilde{B},w) = \widetilde{B}^\top P\widetilde{B}-\widetilde{B}^\top P{\Gamma} P\widetilde{B}
\label{valeurJ}
\end{equation}
with
{\setlength\arraycolsep{0.5pt}\begin{equation}
\begin{array}{rllll}\nonumber
\Gamma := & \sum_{j=0}^{N-1} G^{N-j-1} B_{di,T} R_{d,T}^{-1}B_{di,T}^\top\\
& \times (\1 + PB_{di,T}R_{d,T}^{-1} B_{di,T}^\top )^{-1}(G^\top )^{N-j-1} \\
G := & (\1 + B_{di,T}R_{d,T}^{-1} B_{di,T}^\top P)^{-1}(A_{d,T}-B_{di,T} R_{d,T}^{-1} S_{d,T}^\top ).
\end{array}
\end{equation}}\hfill $\Box$
\end{thm}


\begin{rem}\label{rem:preview}
In Theorem~\ref{LQ-theorem-preview}, we consider a disturbance $w$ made of a single impulse at time $t=NT$. Extending the results of Theorem \ref{LQ-theorem-preview} to multiple simultaneous impulsive inputs can be done as follows. Given $r\in\N_{>0}$ impulses, it suffices to consider $w(t) = c_i\delta(t-NT)$ for any $t\geq 0$, where $c_i$, $i\in\{1, \ldots, r\}$, are the columns of the identity matrix of $\R^{r\times r}$ and compute the square root of the sum of integral squares of the resulting outputs as advocated in \cite{CF91} for continuous-time linear time-invariant systems. Moreover, works as e.g., \cite{KS91,BP91a,CF91} have shown that to connect a continuous-time plant with a sampled controller, as done in this paper, leads to a periodically time-varying closed loop. Consequently, the impulsive input $w$ can occur at any  time, and not only at $NT$. This observation has motivated the introduction of a novel measure for the H$_2$-norm in sampled-data systems, which can be seen as a generalization of the classical H$_2$-norm in \cite{KS91,BP91a,CF91}. The presented results can be extended to fit this framework, we plan to do so in future work.\hfill $\Box$
\end{rem}


\subsection{Controller implementation}
Before we present numerical illustrations of the results of Section \ref{subsect:preview}, we briefly elaborate on how to (approximately) generate the impulsive input $u_{i}$ in (\ref{eq:controlmri}) in practice. The idea consists in applying a constant control during a short interval of time $[0,\alpha_T(\varepsilon)]$, with $\alpha_T(\varepsilon) := \varepsilon T$ for $\varepsilon\in(0,1)$, that is,
\begin{equation}\label{eq:impulse-approximation}
u_i(t) = \begin{cases}
\frac{u_{i_k}}{\alpha_T(\varepsilon)} & \forall t \in \big[t_k, t_k+\alpha_T(\varepsilon) \big) \\
0 & \forall t \in \big[ t_k +\alpha_T(\varepsilon) , t_{k+1} \big).
\end{cases}
\end{equation}
In this case, system (\ref{eq:systemdis}) becomes
\begin{equation}
\begin{array}{ccl}
x_{k+1} = A_{d,T} x_k + B_{d,T} u_{c_k} + B_{i_a,T}(\varepsilon) u_{i_k}
\end{array}
\end{equation}
with $A_{d,T}$, $B_{d,T}$ in (\ref{eq:sampled-matrices}) and $B_{i_a,T}(\varepsilon) := \frac{1}{\alpha_T(\varepsilon)} \int_{0}^{\alpha_T(\varepsilon)} e^{A(T-\tau)} d\tau B$.
When $\varepsilon\rightarrow0$, we recover the MRI sampled-data model (\ref{eq:systemdis}) as $\lim_{\varepsilon\to0}B_{i_a,T}(\varepsilon)=\lim_{\varepsilon\to0}\frac{1}{\alpha_T(\varepsilon)} \int_{0}^{\alpha_T(\varepsilon)} e^{A(T-\tau)} d\tau B=e^{AT} B=B_{i,T}$ where $B_{i,T}$ is defined in~(\ref{eq:sampled-matrices}).
It would be of interest in future work to investigate the impact of the approximation errors due to~(\ref{eq:impulse-approximation}) on the performance of the closed-loop system.

%

\section{Examples}\label{sect:example}

\subsection{Example 1 in \cite{Souza2013}}\label{subsect:ex1-souza}

\begin{figure}[htb]
\begin{center}
\includegraphics[width=0.45\textwidth]{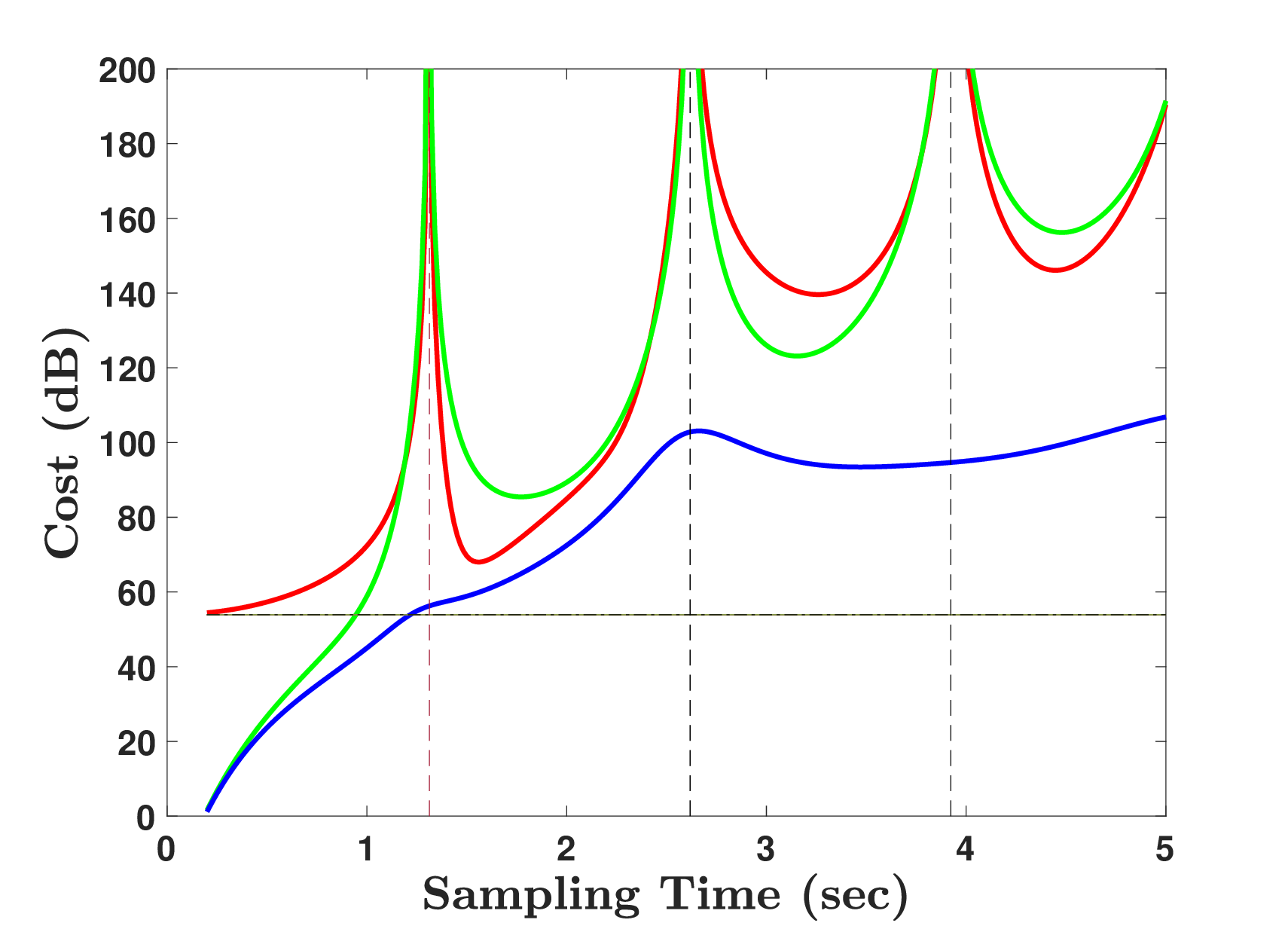}
\caption{\label{fig:mixregimp}Optimal costs in Section \ref{subsect:ex1-souza}: regular LQR  control (red); impulse LQR  control (green); MRI-LQR  control (blue).}
\end{center}
\end{figure}

We first illustrate the results of Sections \ref{sect:controllability} and \ref{subsect:lq} on \cite[Example~1]{Souza2013}, which corresponds to $A,B$ in (\ref{eq:modAB}). As mentioned in Section \ref{sect:controllability}, this example exhibits pathological sampling periods at $T = \dfrac{2\ell\pi}{\sqrt{23}}$ ($\ell = 1, 2, \ldots$). We set
$$
\widetilde{B} = \left[\begin{array}{c} 1 \\ 1 \end{array}\right], \quad 
Q = \left[\begin{array}{cc} 1 & 0\\ 0 &0 \end{array}\right], \quad
R_c = R_i = 1.
$$
In Figure~\ref{fig:mixregimp}, we represent the optimal cost, in a logarithmic scale, evaluated at $\widetilde B$ with sampling periods $T$ ranging from $0.2$ to $5~\text{sec}$ for three distinct scenarios: (i) LQR  regular control ($B_{i,T}=0$ in red); (ii) LQR  pure impulsive control ($B_{d,T}=0$ in green); and (iii) MRI-LQR control (in blue). In configurations (i) and (ii), the corresponding costs blow up at pathological sampling periods. 
This does not occur with the MRI sampled-data model (configuration (iii)), as it is controllable for any $T$. As a result, MRI-LQR control significantly outperforms both regular LQR  and impulsive LQR  in terms of the obtained optimal costs, even when $T$ is not pathological as shown in Figure~\ref{fig:mixregimp}.


\begin{figure}[htb]
\begin{center}
\includegraphics[width=0.45\textwidth]{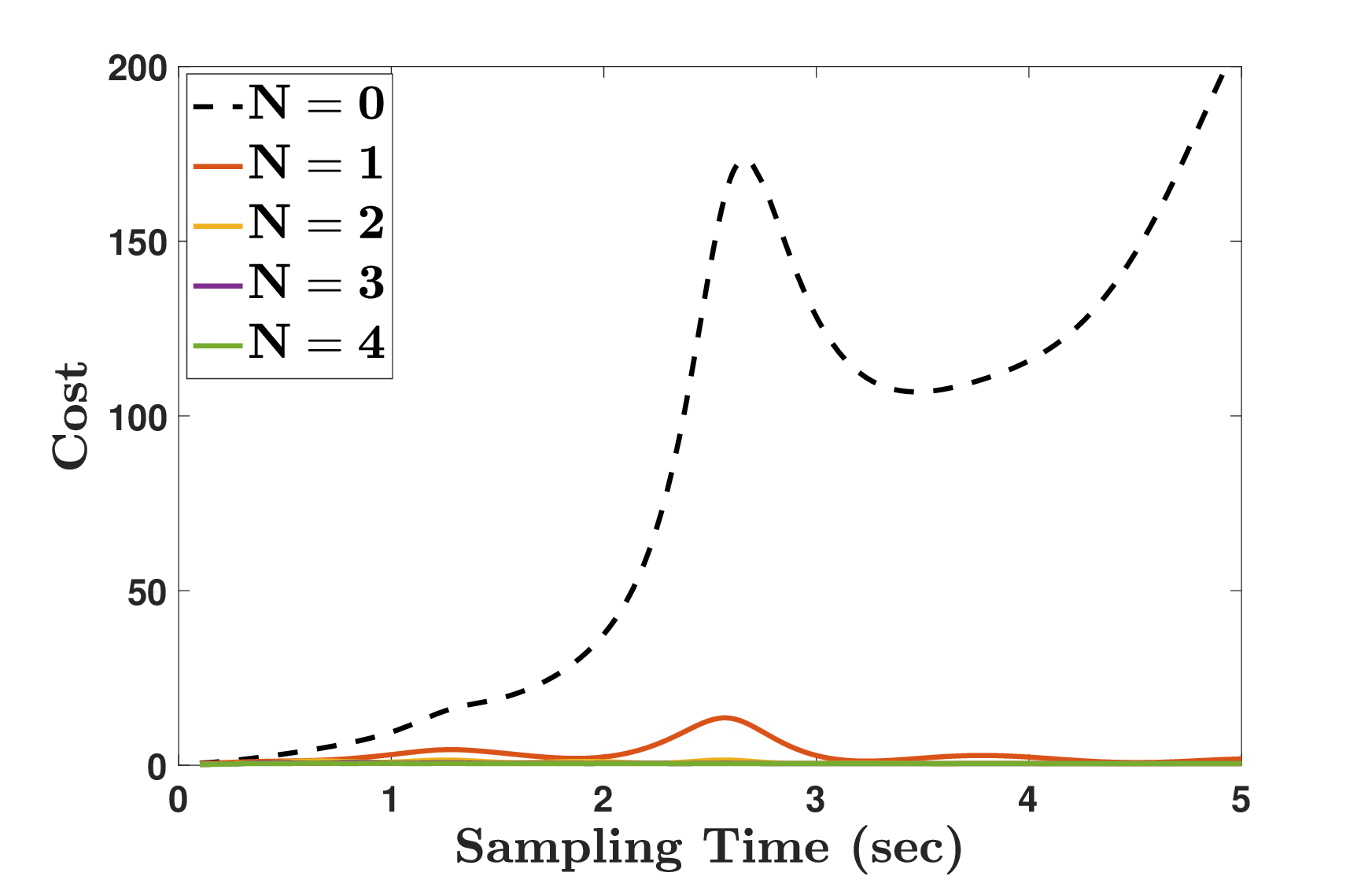}
\caption{\label{fig:withwithoutpreview}Optimal MRI cost in Section \ref{subsect:ex1-souza}: with and without preview.}
\end{center}
 \end{figure}

We then consider preview MRI-LQR control as advocated in Section \ref{subsect:preview}. The outcomes depicted in Figure~\ref{fig:withwithoutpreview} stem from the application of MRI-LQR control with preview, where the external signal $w$ is accessible to the controller $N$ sample time units in advance. The cost is plotted, in a linear scale, versus the sampling period, considering various values of the preview constant $N$ (ranging from 0 to 4) with $N=0$ being the MRI-LQR control without preview. Notably, even with a modest preview value of $N=1$, the incurred cost is consistently lower than that obtained without any preview, as indicated by the dashed line in the plot.



%
%
%

\subsection{Optimal insulin infusion}

We also apply the approach to the optimal insulin infusion of \cite{Good2019} where the primary objective is to minimize the peak of the blood glucose level (BGL) resulting from a food impulse; ignoring the positivity constraint on the input. 
The problem is written in the state space form (\ref{eq:system-w}) with $A = \text{diag}(-0.0167, -0.01, -0.0083, -0.0143, -0.0091, -0.008)$,
$B =\begin{bmatrix} 15 & -75 & 60 & 0 & 0 & 0 \end{bmatrix}^\top$,
$\widetilde{B} =\begin{bmatrix} 0 & 0 & 0 & 1.5909 & -9.1667 & 7.5758\end{bmatrix}^\top$,
$Q = \tilde{C}^\top\tilde{C}$ with
$\tilde{C} =\begin{bmatrix}-1 & -1 & -1 & 1 & 1 & 1\end{bmatrix}$,
$R_c = 1$, $R_i=1$. 
We adopt a sampling period of $T=20$ minutes. 
The obtained MRI-LQR control gains are
$K_c = \left[-0.0962 \, -\!0.1237 \, -\!0.1318 \,\, 0.1052 \,\, 0.1280 \,\, 0.1335 \right]$,
$K_i = \left[-0.0620\, -\!0.0724 \,  -\!0.0752  \,\,  0.0655  \,\,  0.0739  \,\,  0.0758\right]$.
For comparison purposes, we also consider the situation where $R_c=2500$ and $R_i=1$.
This leads to the gains
$\widetilde{K}_c = \left[ -0.0009  \, -\!0.0015 \,  -\!0.0017 \,\,   0.0010 \,\,  0.0016 \,\,  0.0017 \right]$,
$\widetilde{K}_i = \left[ -0.2780 \,  -\!0.4051 \,  -\!0.4464 \,\,  0.3173 \,\,  0.4271 \,\,  0.4552\right]$. 

\begin{figure}[htb]
\begin{center}
\includegraphics[width=0.5\textwidth]{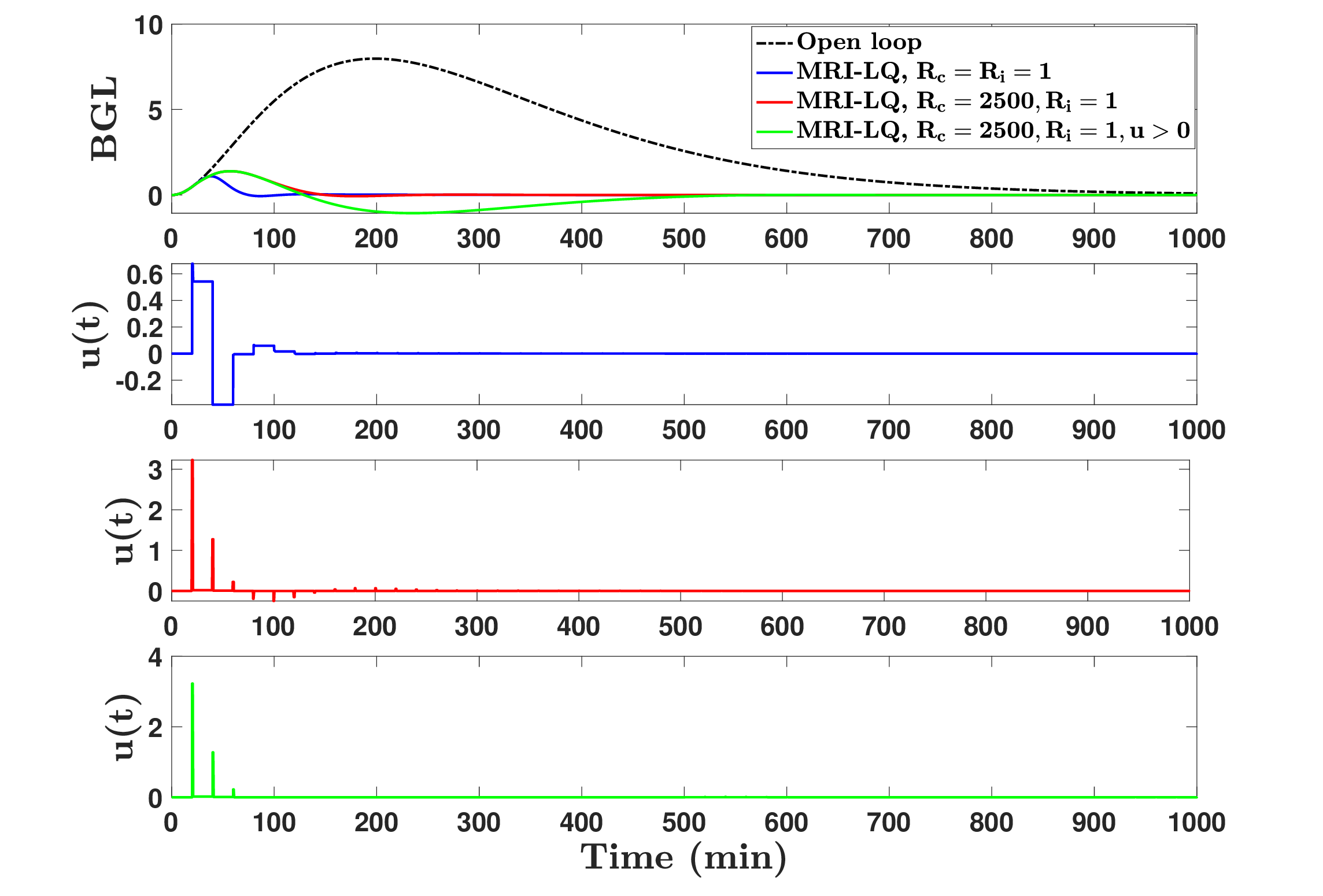}
\caption{\label{fig:insulina}Insulin infusion example: BGL response in open-loop and with different MRI-LQR strategies together with the corresponding control inputs.}
\end{center}
\end{figure}

The results  in Figure~\ref{fig:insulina} depict a scenario where an impulsive disturbance corresponding to a meal intake of $60g$ occurs  at time $0$. 
The open-loop response, in dashed dark line, shows a significant peak excursion in the BGL. 
In contrast, three closed-loop responses are plotted in blue, red, and green, effectively mitigating this peak. The blue response is obtained with $K_c$ and $K_i$. The red response, achieved with $\widetilde{K}_c$ and $\widetilde{K}_i$ highlights the impact of increasing the weight on regular control. 
The green response, on the other hand, corresponds to the solutions obtained when applying $u_{c_k}=\max\{0,\widetilde K_c x_k\}$ and $u_{i_k}=\max\{0,\widetilde K_i x_k\}$ for any $k\in\N$. This saturation reflects the constraint ensuring non-negative insulin flow, presented here solely for illustrative purposes as providing theoretical guarantees in this constrained situation falls outside the scope of this paper. The MRI-LQR control strategy has a physiological interpretation for this example: the impulsive component corresponds to a bolus, and the regular term to a basal input, as mentioned in the introduction. We see in Figure~\ref{fig:insulina} that the blue response involves a single bolus application at $T=20$ minutes, aligning with the next sampling period after the disturbance impulse. The red response displays more impulses with a concurrent reduction in the magnitude of the regular component. We can check that its saturated version effectively reduces the BGL peak. 

\begin{figure}[htb]
\begin{center}
\includegraphics[width=0.4\textwidth]{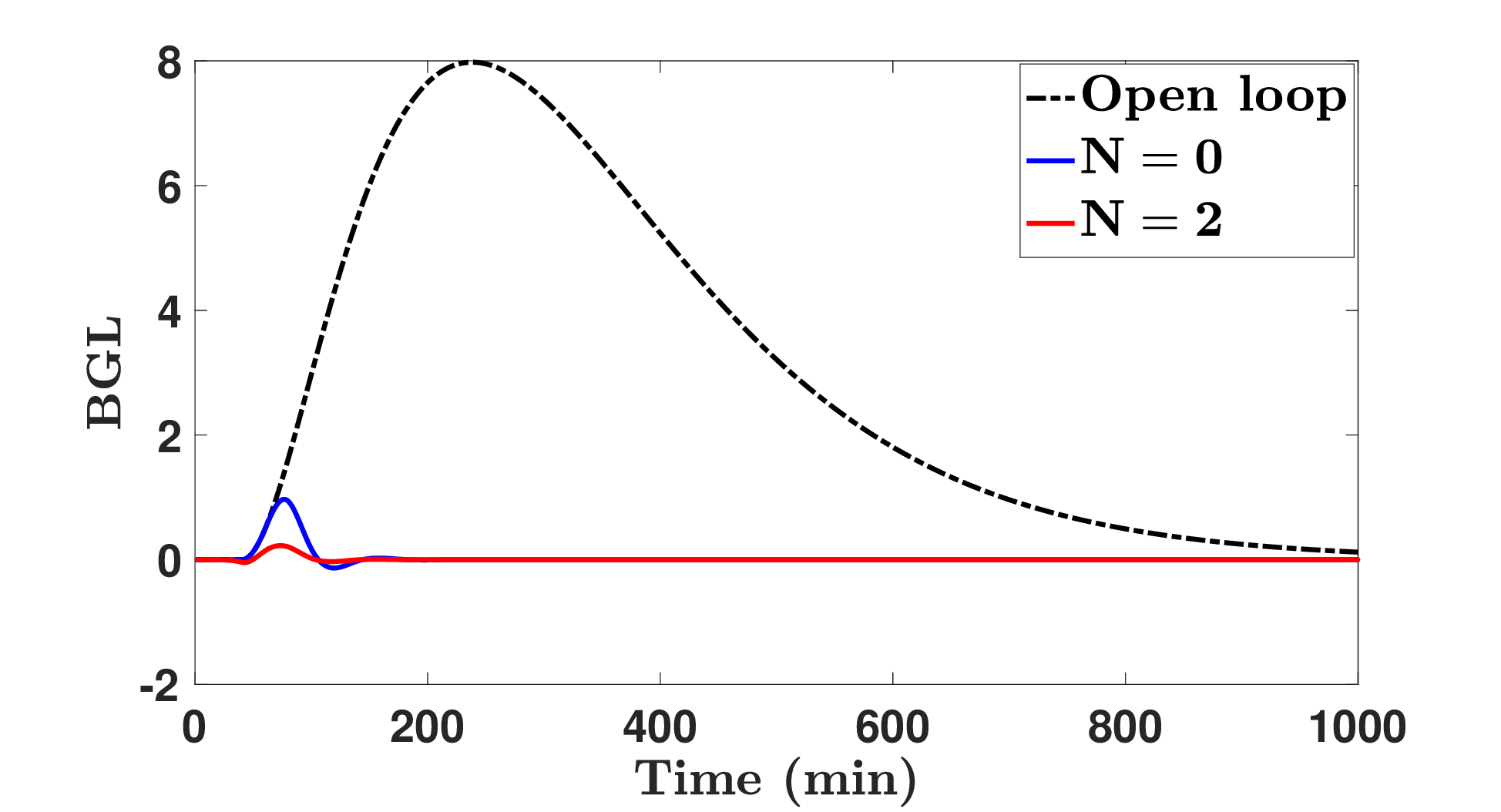}
\caption{\label{fig:insulinapreview}Insulin infusion example: BGL response given by the MRI-LQR strategy with and without preview ($N=2$ and $N=0$, respectively).}
\end{center}
\end{figure}

Finally, we notice the benefit of the MRI-LQR control with preview in Figure~\ref{fig:insulinapreview}. When comparing the MRI-LQR strategy without preview (depicted in blue) to the closed-loop response achieved with a preview of $N=2$ (illustrated by the red curve), the latter result shows a desirable significantly reduced peak in the blood glucose level. Indeed, the MRI-LQR strategy with preview leads to a significant peak reduction in the blood glucose level. This aligns with medical advice advocating for insulin injection prior to meals, which has been shown to significantly enhance post-meal control~\cite{Slattery18}.

\section{Conclusion}\label{sect:conclusion}

We have illustrated the benefits of mixing regular and impulsive inputs for the sampled-data control of linear time-invariant systems. In particular, adding an impulsive term to a regular control can help to reduce the set of pathological sampling periods. We have also derived (preview) LQR  control in this context and shown via two examples that significant performance can be obtained. 

Various extensions can be envisioned  for the sampled-data control of linear time-invariant systems, as already hinted in Remark \ref{rem:preview}, but more generally for other classes of dynamical systems for which a comprehensive theory of optimal MRI control is missing.




\section*{Appendix: Proof of Theorem~\ref{LQ-theorem-preview}}

Let $\widetilde B\in\R^{n}$, $N\in\N_{>0}$, $T\in\Rlp$ be non-pathological for system (\ref{eq:systemdis}) and $w$ be as in Theorem \ref{LQ-theorem-preview}. In view of the explanations after (\ref{CostH2}), we write $\widetilde J$ as, for $u$ given by (\ref{eq:controlmri}),
\begin{equation}
\widetilde{J}(u,w) = \displaystyle \int_0^{NT} z(t)^\top z(t) dt + \int_{NT}^{\infty} z(t)^\top z(t) dt.
\end{equation}
By Lemma \ref{lem:equality-cost}, for any $u$ as in (\ref{eq:controlmri}), we have
{\setlength\arraycolsep{0.5pt}\begin{equation}
\begin{array}{rllll}
J_{d,T}(\widetilde B,\mathbf{v}) & := & \displaystyle\int_{NT}^{\infty} z (t)^\top z(t) dt\\
& = & \displaystyle\sum_{k=0}^{\infty} \Big( x_k^\top Q_{d,T} x_k + 2 x_k^\top S_{d,T} v_k + v_k^\top R_{d,T} v_k \Big),
\end{array}
\end{equation}}
\hspace{-0.15cm}with $\mathbf{v}=(v_k)_{k\in\N}$, $v_k=\left[\begin{matrix}u_{c_k}\\u_{i_k}\end{matrix}\right]$ and $x_k$ is the solution to (\ref{eq:systemdis}) with initial condition $\widetilde B$. We also denote for the sake of convenience
{\setlength\arraycolsep{2pt}\begin{equation}
J_{c,T}(u,w) := \displaystyle \int_0^{NT} z^\top (t)z(t) dt.
\end{equation}}
On $[NT, \infty)$, the problem reduces to the infinite-horizon MRI-LQR case, that is (\ref{eq:systemdis}) with the initial condition $x_N+\widetilde{B}$ and cost (\ref{equalityJ}). It follows that the optimal solution is the state-feedback given by (\ref{u2et}). Consequently, the optimal cost is equal to $(x_N+\widetilde{B})^\top P(x_N+\widetilde{B})$ with $P=P^\top>0$ solution to~(\ref{DARE1}). Now, we focus on the interval $[0,NT]$ and $J_{c,T}(u,w)$. The dynamical system to be considered is (\ref{eq:system-w}) with $w=0$ within this time interval. Similar arguments as in the proof of Lemma \ref{lem:equality-cost} lead to
\begin{equation}
		J_{c,T}(u,w) = \sum_{k=0}^{N-1} \Big( x_k^\top Q_{d,T} x_k + 2x_k^\top S_{d,T}v_{k} + v_{k}^\top R_{d,T}v_{k} \Big)
		\label{Jdiscfini} 
\end{equation}
with $v_k=\left[\begin{matrix}u_{c_k}\\u_{i_k}\end{matrix}\right]$, $u$ as in (\ref{eq:controlmri}) and $x_k$ is the solution to (\ref{eq:systemdis}). As a consequence, we have
\begin{equation}
\begin{array}{r@{\,}l}
\widetilde{J}(u,w) &= \displaystyle\sum_{k=0}^{N-1} \Big( x_k^\top Q_{d,T} x_k + 2x_k^\top S_{d,T}v_{k} + v_{k}^\top R_{d,T}v_{k} \Big) \\
& + (x_N+\widetilde{B})^\top P(x_N+\widetilde{B})
\end{array}
\end{equation}
To derive the inputs minimizing the above cost along solutions to (\ref{eq:systemdis}), we apply the conventional minimization method using Lagrange multipliers and consider $\frac{1}{2}J$ for the sake of convenience. We obtain
\begin{align}
	x_{k+1} &= A_{d,T}x_k + B_{di,T}v_{k} \label{xk1} \\
	\mu _{k} &= A_{d,T}^\top \mu_{k+1} + Q_{d,T}x_k +S_{d,T} v_{k} \label{lamk} \\
	v_{k} &= -R_{d,T}^{-1}B_{di,T}^\top \mu _{k+1}-R_{d,T}^{-1}S_{d,T}^\top x_k \label{convk}\\
	\mu _N &= P(x_N+\widetilde{B}) \label{lamda}
\end{align}
with $\mu_k$ the so-called Lagrange multiplier or adjoint vector.
Let
\begin{equation}
	q_k := Px_k-\mu_k.
	\label{qPlam}
\end{equation}
Using this notation and replacing $v_k$ in (\ref{xk1}) and (\ref{lamk}) by its expression (\ref{convk}), we obtain
\begin{equation}
\begin{array}{r@{\,}c@{\,}l}
		x_{k+1} & =& (\1 + B_{di,T}R_{d,T}^{-1} B_{di,T}^\top P)^{-1}\big( (A_{d,T} \\
    & &-B_{di,T} R_{d,T}^{-1} S_{d,T}^\top )x_k 
		+ B_{di,T} R_{d,T}^{-1} B_{di,T}^\top q_{k+1}\big) \label{xk1q}
	\end{array}
\end{equation}
and
\begin{equation}
	\begin{array}{r@{\,}c@{\,}l}	
		q _{k} & = & Px_k-(A_{d,T}-B_{di,T} R_{d,T}^{-1} S_{d,T}^\top )^\top \mu_{k+1} \\
		& & -(Q_{d,T}-S_{d,T} R_{d,T}^{-1} S_{d,T}^\top )x_k.
		\label{Pqn}
	\end{array}
\end{equation}
Using $\mu _{k+1} = Px_{k+1}-q_{k+1}$ with $x_{k+1}$ given by (\ref{xk1q}), we deduce, after some algebraic manipulations,
$$
\begin{array}{r@{\,}c@{\,}l}
	q _{k} & = & (A_{d,T}-B_{di,T} R_{d,T}^{-1} S_{d,T}^\top )^\top q_{k+1} \\
& & -(A_{d,T}-B_{di,T} R_{d,T}^{-1} S_{d,T}^\top )^\top P\\
& & \times(\1 + B_{di,T}R_{d,T}^{-1} B_{di,T}^\top P)^{-1} B_{di,T} R_{d,T}^{-1} B_{di,T}^\top q_{k+1} \\
& & + (P-Q_{d,T} + S_{d,T} R_{d,T}^{-1} S_{d,T}^\top ) x_k \\
& & -(A_{d,T}-B_{di,T} R_{d,T}^{-1} S_{d,T}^\top )^\top P\\
& & \times (\1 + B_{di,T}R_{d,T}^{-1} B_{di,T}^\top P)^{-1} \\
& & \times(A_{d,T}-B_{di,T} R_{d,T}^{-1} S_{d,T}^\top )x_k.
\end{array}
$$
As $P$ is solution to (\ref{DARE1}), which is equivalent to $(A_{d,T}-B_{di,T}R_{d,T}^{-1}S_{d,T}^\top )^\top P(\1+B_{di,T}R_{d,T}^{-1}B_{di,T}^\top P)^{-1}(A_{d,T}
   -B_{di,T}R_{d,T}^{-1}S_{d,T}^\top )
			+Q_{d,T}-S_{d,T}R_{d,T}^{-1}S_{d,T}^\top = P$, 
and using the matrix inversion lemma, we obtain
\begin{equation}
	q_k = G^\top q_{k+1},
\end{equation}
with
$G = (\1 + B_{di,T}R_{d,T}^{-1} B_{di,T}^\top P)^{-1}(A_{d,T}-B_{di,T} R_{d,T}^{-1} S_{d,T}^\top )$.
As a consequence, we have
$$
q_k = (G^\top )^{-k}q_0, \quad q_N = (G^\top )^{-N}q_0, \quad q_0 = (G^\top )^{N}q_N.
$$
Notice also, from (\ref{lamda}) and (\ref{qPlam}), that	$q_N = -P\widetilde{B}$.
Hence,
\begin{equation}
	q_k = -(G^\top )^{N-k}P\widetilde{B}.
	\label{equaqk}
\end{equation}
This leads to
$\mu_k = Px_k+(G^\top )^{N-k}P\widetilde{B}$
and
$$
{\setlength\arraycolsep{0.5pt}\begin{array}{rllll}
	v_{k} & = & -R_{d,T}^{-1}B_{di,T}^\top P(A_{d,T}x_k+B_{di,T}v_{k} )- R_{d,T}^{-1}S_{d,T}^\top x_k \\
	& & -R_{d,T}^{-1}B_{di,T}^\top (G^\top )^{N-k-1}P\widetilde{B}\\
	& = & -(\1 + R_{d,T}^{-1} B_{di,T}^\top PB_{di,T})^{-1}R_{d,T}^{-1}B_{di,T}^\top PA_{d,T}x_k \\
	& & -(\1 + R_{d,T}^{-1} B_{di,T}^\top PB_{di,T})^{-1}R_{d,T}^{-1}S_{d,T}^\top x_k \\
	& & - (\1 + R_{d,T}^{-1} B_{di,T}^\top PB_{di,T})^{-1}R_{d,T}^{-1}B_{di,T}^\top \\
 & & \times(G^\top )^{N-k-1}P\widetilde{B}.
\end{array}}
$$
\hspace{-0.15cm}Finally, we obtain
\begin{equation}
	\begin{array}{l}\nonumber
		v_{k}^* = -(R_{d,T}+B_{di,T}^\top PB_{di,T})^{-1}(B_{di,T}^\top PA_{d,T}+S_{d,T}^\top )x_k \\
		\quad \quad \quad      -(R_{d,T}+B_{di,T}^\top PB_{di,T})^{-1}B_{di,T}^\top (G^\top )^{N-k-1}P\widetilde{B}.
	\end{array}
\end{equation}

To conclude the proof, it remains to compute the explicit cost formula (\ref{valeurJ}).
To this end, notice that the criterion given by~(\ref{Jdiscfini}) can be written as
{\setlength\arraycolsep{0.5pt}\begin{equation}
\begin{array}{rllll}
\widetilde{J}(u,w) & = & (x_N+\widetilde{B})^\top P(x_N+\widetilde{B}) \\
& & \displaystyle + 
\sum_{k=0}^{N-1} \Big( x_k^\top (Q_{d,T}-S_{d,T}R_{d,T}^{-1}S_{d,T}^\top ) x_k \\
& & + (v_{k}+ R_{d,T}^{-1}S_{d,T}^\top x_k)^\top R_{d,T}(v_{k}+ R_{d,T}^{-1}S_{d,T}^\top x_k) \Big).
\end{array}\nonumber
\end{equation}}
\hspace{-0.15cm}From (\ref{Pqn}), we deduce
$$
\begin{array}{r@{\,}c@{\,}l}
	x_k^\top Px_k &=& x_k^\top (Q_{d,T}-S_{d,T}R_{d,T}^{-1}S_{d,T}^\top )x_k\\
	&&+x_k^\top (A_{d,T}-B_{di,T}R_{d,T}^{-1}S_{d,T}^\top )^\top Px_{k+1}\\
	&& -x_k^\top (A_{d,T}-B_{di,T}R_{d,T}^{-1}S_{d,T}^\top )^\top q_{k+1}+x_k^\top q_k.
\end{array}
$$
By (\ref{xk1q}), 
$$\begin{array}{r@{\,}c@{\,}l}
	x_k^\top Px_k &=& x_k^\top q_k + x_k^\top (Q_{d,T}-S_{d,T}R_{d,T}^{-1}S_{d,T}^\top )x_k \\
&&-\Big( x_{k+1}^\top (\1 + B_{di,T}R_{d,T}^{-1} B_{di,T}^\top P)\\
&& -q_{k+1}^\top B_{di,T}R_{d,T}^{-1} B_{di,T}^\top \Big)q_{k+1} \\
&&+\Big( x_{k+1}^\top (\1 + B_{di,T}R_{d,T}^{-1} B_{di,T}^\top P)\\
&& -q_{k+1}^\top B_{di,T}R_{d,T}^{-1} B_{di,T}^\top \Big)Px_{k+1},
\end{array}
$$
which is equivalent to
$$\begin{array}{l}
    x_k^\top (Q_{d,T}-S_{d,T}R_{d,T}^{-1}S_{d,T}^\top )x_k\\
    \hspace{5mm}=-x_{k+1}^\top Px_{k+1} + x_k^\top Px_k \\
	\hspace{5mm}- (Px_{k+1}-q_{k+1})^\top B_{di,T}R_{d,T}^{-1}B_{di,T}^\top (Px_{k+1}-q_{k+1})\\
	\hspace{5mm}+x_{k+1}^\top q_{k+1}-x_k^\top q_k.
\end{array}$$
Combining (\ref{convk}) and (\ref{qPlam}) give 
$$
v_{k}+ R_{d,T}^{-1}S_{d,T}^\top x_k = -R_{d,T}^{-1}B_{di,T}^\top (Px_{k+1}-q_{k+1})
$$
and therefore
$$
\begin{array}{l}
	x_k^\top (Q_{d,T}-S_{d,T}R_{d,T}^{-1}S_{d,T}^\top )x_k \\
 + (v_{k}+ R_{d,T}^{-1}S_{d,T}^\top x_k)^\top R_{d,T}(v_{k}+ R_{d,T}^{-1}S_{d,T}^\top x_k) \\ = x_k^\top Px_k-x_{k+1}^\top Px_{k+1} + x_{k+1}^\top q_{k+1}-x_k^\top q_k.
\end{array}
$$
As a consequence
\begin{equation}
\begin{array}{r@{\,}c@{\,}l}
\widetilde{J}(u,w) & = & (x_N+\widetilde{B})^\top P(x_N+\widetilde{B}) 	+ x_0^\top Px_0-x_N^\top Px_N\\
& & +\sum_{k=0}^{N-1}(x_{k+1}^\top q_{k+1}-x_k^\top q_k).
\end{array}
\end{equation}
Using $x_0=0$ and $q_N = -P\widetilde{B}$, one gets
$$
\begin{array}{r@{\,}c@{\,}l}
	\widetilde{J}(u,w) &=& (x_N+\widetilde{B})^\top P(x_N+\widetilde{B}) + x_0^\top Px_0 \\
 & & -x_N^\top Px_N +x_{N}^\top q_{N}-x_0^\top q_0\\
	&=& B_2^\top P\widetilde{B}+2x_N^\top P\widetilde{B}+x_N^\top q_N\\
	&=& x_N^\top P\widetilde{B} + \widetilde{B}^\top P\widetilde{B}.
\end{array}
$$
To obtain (\ref{valeurJ}), we are left with computing $x_N$. From (\ref{xk1q}) and (\ref{equaqk}), we deduce
\begin{equation}
\begin{array}{r@{\,}c@{\,}l}
x_{k+1} & = & Gx_k + (\1 + B_{di,T}R_{d,T}^{-1} B_{di,T}^\top P)^{-1} B_{di,T} \\
& & R_{d,T}^{-1}B_{di,T}^\top q_{k+1}
\end{array}
\end{equation}
and
$$
\begin{array}{r@{\,}c@{\,}l}
	x_k &=& G^kx_0- \sum_{j=0}^{k-1} G^{k-j-1} (\1 + B_{di,T}R_{d,T}^{-1} B_{di,T}^\top P)^{-1}\\
	&& \times B_{di,T} R_{d,T}^{-1}B_{di,T}^\top (G^\top )^{N-j-1} P\widetilde{B}.
\end{array}
$$
At $k=N$, we have
$$
\begin{array}{r@{\,}c@{\,}l}
	x_N &=& -\sum_{j=0}^{N-1} G^{N-j-1} (\1 + B_{di,T}R_{d,T}^{-1} B_{di,T}^\top P)^{-1}\\
	& & \times B_{di,T} R_{d,T}^{-1}B_{di,T}^\top (G^\top )^{N-j-1} P\widetilde{B}
\end{array}
$$
and (\ref{valeurJ}) follows. This concludes the proof.

\bibliographystyle{IEEEtran}
\bibliography{MixRI}

\end{document}